\documentclass{webofc}
\usepackage[varg]{txfonts}   
\newcommand{\di}{{\rm d}}

\def\wT{{\widehat T}}
\def\wj{{\widehat j}}
\def\wJ{{\widehat J}}

\def\wP{{\widehat P}}

\def\wQ{{\widehat Q}}
\def\wspt{{\widehat{\cal S}}}

\def\wrho{{\widehat{\rho}}}
\def\wrhol{{\widehat{\rho}_{\rm LE}}}

\newcommand{\tr}{{\rm tr}}

\newcommand{\omegav}{\boldsymbol{\omega}}

\newcommand{\snn}{\sqrt{s_{\rm NN}}}
\newcommand{\be}{\begin{equation}}
\newcommand{\ee}{\end{equation}}                                                                               
\def\bea{\begin{eqnarray}}
\def\eea{\end{eqnarray}}

\begin{document}
\title{Polarization in relativistic heavy ion collisions: a theoretical perspective}
%
%

\author{\firstname{Francesco} \lastname{Becattini}\inst{1}\thanks{\email{
becattini@fi.infn.it}}}

\institute{Dipartimento di Fisica e Astronomia, University of Florence, Via G. Sansone 1,
I-50019, Sesto Fiorentino (Firenze), Italy}

\abstract{We review the theoretical framework for the calculation of particle 
polarization in relativistic heavy ion collisions within the hydrodynamical model. 
The covariant decomposition of the mean spin vector is presented and open theoretical 
issues addressed.}
\maketitle
\section{Introduction}\label{intro}

Global polarization of $\Lambda$ hyperons in semi-peripheral relativistic heavy-ion 
collisions has been recently observed by the STAR experiment over a center-of-mass 
energy range between 7.7 and 200 GeV \cite{STAR:2017ckg}. This finding 
confirms quantitative predictions based on local thermodynamic equilibrium of 
spin degrees of freedom \cite{becaspin} which provides a relation between 
polarization and relativistic vorticity, and it agrees quantitatively with the 
hydrodynamic model calculations to a very good degree of accuracy. In the hydrodynamic 
framework, the distinctive feature of polarization is its proportionality to the 
gradients of the combined temperature and velocity fields, so that its measurement 
is a stringent test of the hydrodynamic picture which is distinct and complementary 
to momentum spectra. Moreover, hydrodynamics and local thermodynamic equilibrium
predict that polarization must be the same for particles and antiparticles, that
is a C-even effect, in agreement with the observations. Conversely, if electromagnetic
or other C-odd fields were responsible for this phenomenon, polarization would be
{\em opposite} for particles and antiparticles. 

The experimental efforts, thus far, focussed on the search of the average global 
polarization of $\Lambda$ hyperons along the direction of the angular momentum of 
the plasma. This measurement requires the identification of the reaction plane in 
peripheral collision as well as its orientation, that is the direction of the total 
angular momentum vector ${\bf J}$. 
The average global polarization along ${\bf J}$ is also found to decrease rapidly 
as a function of center-of-mass energy \cite{STAR:2017ckg}, from few percent at 
$\snn = {\cal O}(10)$ GeV to few permille at $\snn = {\cal O}(100)$ GeV. In the TeV 
energy range, at the LHC, the global polarization along ${\bf J}$ is not seen 
\cite{alice} as it is most likely beyond experimental sensitivity.

\section{The theoretical approaches to global polarization: a summary}

Particles produced in relativistic heavy ion collisions are expected to be polarized 
in peripheral collisions because of angular momentum conservation. At finite impact 
parameter, the Quark Gluon Plasma (QGP) has a finite angular momentum perpendicular 
to the reaction plane and some fraction thereof may be converted into spin of final 
state hadrons. Therefore, measured particles may show a finite mean {\em global} 
polarization along the angular momentum direction. 

Early estimates of this effect \cite{xnwang1} were based on the general idea that 
polarized quarks in the QGP stage of the production process would eventually give 
rise to polarized hadrons, making it possible to predict qualitative features of 
the final hadrons' polarization. It was then proposed \cite{becapicc1,becapicc2} 
that polarization can be calculated assuming that the spin degrees of freedom are 
at local thermodynamical equilibrium at the hadronization stage in much the same 
way as the momentum degrees of freedom. In other words, polarization can be predicted 
by extending the familiar Cooper-Frye formula to particles with spin. A specific 
derivation was presented in refs.~\cite{becapicc1,becaspin} where it was pointed 
out that the hydrodynamical quantity steering the polarization is the {\em thermal 
vorticity}, that is (minus) the antisymmetric part of the gradient of the four-temperature 
field $\beta = (1/T) u$ where $T$ is the proper temperature and $u$ the hydrodynamic 
four-velocity:
\be\label{thvort}
   \varpi_{\mu\nu} = -\frac{1}{2} \left( \partial_\mu \beta_\nu - \partial_\nu 
   \beta_\mu \right)
\ee
Particularly, the first-order expansion of the polarization in terms
of thermal vorticity was obtained in ref.~\cite{becaspin} for hadrons with spin $1/2$
(lately recovered with a different method in ref.~\cite{xnwang2}), yet its extension 
to higher spins could be derived from the corresponding global equilibrium 
expression \cite{becalisa}. 

This theoretical work made it possible to make definite quantitative predictions 
of global $\Lambda$ polarization in nuclear collisions from hydrodynamic calculations, 
with a resulting mean value ranging from some permille to some percent~\cite{becacsernai,
beca2015,xnwang3}, with an apparently strong dependence on the initial conditions, 
particularly on the initial longitudinal velocity field. Calculations of vorticity 
in relativistic heavy ion collisions - which could be then turned into a polarization map -
were also recently presented in ref.~\cite{liao}. To complete the theoretical overview 
on the subject, it should be pointed out that different approaches, as well as 
additional mechanisms, to the $\Lambda$ polarization in relativistic nuclear collisions 
were proposed in refs.~\cite{ayala,celso,teryaev, aristova}.

\subsection*{Notation}

In this paper we use the natural units, with $\hbar=c=K=1$.\\ 
The Minkowskian metric tensor is ${\rm diag}(1,-1,-1,-1)$; for the Levi-Civita
symbol we use the convention $\epsilon^{0123}=1$.\\  
Operators in Hilbert space will be denoted by a large upper hat, e.g. $\wT$ while unit 
vectors with a small upper hat, e.g. $\hat v$.

\section{Hydrodynamics and the local equiibrium density operator}

The proper relativistic extension of the spin concept, for massive particles, requires 
the introduction of a spin four-vector operator. This is defined as follows:
\begin{equation}
 \widehat{S}^\mu = -\frac{1}{2m} 
\epsilon^{\mu\nu\rho\lambda} \widehat{J}_{\nu\rho} \widehat{p}_\lambda
\end{equation}
where $\widehat{J}$ and $\widehat{p}$ are the angular momentum operator and four-momentum 
operator of a single particle. As it can be easily shown, the spin four-vector 
operator commutes with the four-momentum operator (hence it is a compatible observable) 
and it is space-like on free particle states as it is orthogonal to the four-momentum:
\be
  \widehat{S}^\mu \widehat{p}_\mu = 0
\ee
and has thus only three independent components. 

The mean spin $S^\mu$ and polarization $P^\mu$ four-vectors can now be defined as 
the mean values with the suitable density operators, that is:
\be\label{spinv}
 S^\mu = \langle \widehat{S}^\mu \rangle \equiv \tr (\wrho \, \widehat{S}^\mu )
\ee
and
\be\label{polav}
 P^\mu = \langle \widehat{S}^\mu \rangle/ S
\ee

The crucial role in the calculation of (\ref{spinv}) for the fluid produced in 
relativistic heavy ion collisions is played by the density operator. For a system 
at local thermodynamic equilibrium, this reads \cite{betaframe}:
\be\label{gencov}
  \wrhol = (1/Z) \exp \left[- \int_\Sigma \di\Sigma_\mu  \left( \wT^{\mu\nu} \beta_\nu 
- \zeta \wj^\mu \right) \right]
\ee
where $\beta$ is the four-temperature vector, $\wT$ the stress-energy tensor, $\wj$
a conserved current - like the baryon number - and $\zeta=\mu/T$. 

If we wish to calculate the mean value of a local operator $\widehat O(x)$ (such 
as, for instance the stress-energy tensor $\wT$, or the current $\wj$)
\be
  O(x) = \tr ( \wrho \widehat O(x) )
\ee
and if the fields $\beta$,$\zeta$ vary significantly over a distance which is much 
larger than the typical microscopic length (indeed the {\em hydrodynamic limit}), 
then they can be Taylor expanded in the density operator starting from the point 
$x$ where the mean value $O(x)$ is to be calculated. The leading terms in the 
exponent of (\ref{gencov}) then become \cite{betaframe}:
\be\label{densop}
 \wrhol \simeq \frac{1}{Z_{\rm LE}}\exp \left[- \beta_\nu(x) \widehat P^\nu + 
 \xi(x) \widehat Q  - \frac{1}{4} (\partial_\nu \beta_\lambda(x) - \partial_\lambda 
 \beta_\nu(x)) \widehat J_x^{\lambda\nu} + \frac{1}{2}(\partial_\nu \beta_\lambda(x) 
 + \partial_\lambda \beta_\nu(x)) \,\widehat L^{\lambda\nu}_x + \nabla_\lambda \xi(x) 
 \, \widehat d^\lambda_x \right].
\ee
where the last two terms with the shear tensor and the gradient of $\zeta$ are 
dissipative and vanish at equilibrium. The $\nabla_\lambda$ operator stands for:
$$
  \nabla_\lambda = \partial_\lambda - u_\lambda u \cdot \partial
$$
as usual in relativistic hydrodynamics. The term which is responsible for polarization 
is the one involving the generators of the Lorentz group $\widehat J_x$.

The polarization of particles in a fluid was then obtained by means of the spin
tensor $\wspt$ which is the local operator associated to spin density along with
an ansatz about the form of the covariant Wigner function \cite{becaspin}.
The mean spin vector of $1/2$ particles with four-momentum $p$, produced around 
point $x$ at the leading order in the thermal vorticity is then found to be:
\begin{equation}\label{basic}
  S^\mu(x,p)= - \frac{1}{8m} (1-n_F) \epsilon^{\mu\rho\sigma\tau} p_\tau \varpi_{\rho\sigma}
\end{equation}
where $n_F = (1+\exp[\beta(x) \cdot p - \nu(x) Q/T(x)] +1)^{-1}$ is the Fermi-Dirac 
distribution and $\varpi(x)$ is given by eq.~\ref{thvort}. The eq.~(\ref{basic}) 
has been recovered with a different approach in ref.~\cite{xnwang2}. This formula 
is suitable for the situation of relativistic heavy ion collisions, where one deals 
with a local thermodynamic equilibrium hypersurface $\Sigma$ where hydrodynamic 
stage ceases and particle description sets in. 

\section{Covariant decomposition of the spin vector in a relativistic fluid}

To gain insight into the physics of polarization in a relativistic fluid, it is 
very useful to decompose the gradients of the four-temperature vector in the 
eq.~(\ref{basic}). We start off with the seperation of the gradients of the comoving 
temperature and four-velocity field:
$$
 \partial_\mu \beta_\nu = \partial_\mu \left(\frac{1}{T}\right) + \frac{1}{T} 
 \partial_\mu u_\nu 
$$
Then, we can introduce the acceleration and the vorticity vector $\omega^\mu$ with
the usual definitions:
\begin{eqnarray*}
A^\mu &= &u \cdot \partial u^\mu  \\
\omega^\mu &=& \frac{1}{2} \epsilon^{\mu\nu\rho\sigma} \partial_\nu u_\rho u_\sigma
\end{eqnarray*}
The antisymmetric part of the tensor $\partial_\mu u_\nu$ can then be expressed as
a function of $A$ and $\omega$:
$$
  \frac{1}{2}\left( \partial_\nu u_\mu - \partial_\mu u_\nu \right) = \frac{1}{2} 
   \left( A_\mu u_\nu - A_\nu u_\mu \right) + \epsilon_{\mu\nu\rho\sigma} \omega^\rho 
  u^\sigma
$$
therafter plugged into the (\ref{basic}) to give:
\begin{eqnarray}\label{spindeco}
S^\mu(x,p)  &=& \frac{1}{8m} (1-n_F)  \epsilon^{\mu\nu\rho\sigma} p_\sigma
   \nabla_\nu (1/T) u_\rho \\
&+&  \frac{1}{8m} (1-n_F) \; 2 \, \frac{\omega^\mu u \cdot p - u^\mu \omega \cdot p}{T} \\
&-& \frac{1}{8m}(1-n_F)\frac{1}{T} \epsilon^{\mu\nu\rho\sigma} p_\sigma  A_\nu u_\rho
\end{eqnarray}
Hence, polarization stems from three contributions: a term proportional to the 
gradient of temperature, a term proportional to the vorticity $\omega$, and a term
proportional to the acceleration. Further insight into the nature of these terms
can be gained by choosing the particle rest frame, where $p=(m,{\bf 0})$ and restoring
the natural units. The eq.~(\ref{spindeco}) then certifies that the spin in the 
rest frame is proportional to the following combination:
\be\label{restframe}
{\bf S}^*(x,p) \propto \frac{\hbar}{KT^2} \gamma {\bf v} \times \nabla T + 
\frac{\hbar}{KT} \gamma (\omegav - (\omegav \cdot {\bf v}) {\bf v}/c^2) 
+ \frac{\hbar}{KT} \gamma {\bf A} \times {\bf v}/c^2
\ee
where $\gamma = 1/\sqrt{1-v^2/c^2}$ and all three-vectors, including vorticity,
acceleration and velocity, are observed in the particle rest frame. 

The three independent contributions are now well discernible in eq.~(\ref{restframe}). 
The second term scales like $\hbar \omega/KT$ and is the one already known from 
non-relativistic physics, proportional to the vorticity vector seen by the particle 
in its motion amid the fluid, with an additional term vanishing in the non-relativistic 
limit. The third term is a purely relativistic one and scales like $\hbar A/KTc^2$;
it is usually overwhelmingly suppressed, except in heavy ion collisions where the 
acceleration of the plasma is huge ($A \sim 10^{30} g$ at the outset of hydrodynamical
stage). The first term, instead, is a new non-relativistic term \cite{becaspin} and 
applies to situations where the velocity field is not parallel to the temperature 
gradient. For ideal uncharged (thus relativistic) fluids, this term is related 
to the acceleration term because the equations of motion reduce to:
$$
 \nabla_\mu T = T A_\mu/c^2
$$
Therefore, being the QGP a quasi-ideal fluid and almost uncharged at very high 
energy, the first and third term are tightly related. It can be shown that they
contribute non-trivially to the final predicted polarization \cite{becakarp}.

\section{Open theoretical issues}

Specifying the mean spin vector at some point $x$ requires the spin tensor $\wspt$,
what was used in ref.~\cite{becaspin} to obtain the function in eq.~(\ref{basic}). 
However, in relativistic heavy ion collisions, we do not measure spin at some 
spacetime point, but only as a function of particle momentum. In formulae, what we
can do is to measure the mean spin after an integration of (\ref{basic}) over the 
freeze-out or particlization hypersurface $\Sigma$.
\be\label{basic2}
 S^\mu(p)= \frac{\int d\Sigma_\lambda p^\lambda f(x,p) S^\mu(x,p)}{\int d\Sigma_\lambda 
 p^\lambda f(x,p)}
\ee

It is presently an unsettled issue whether a formula like (\ref{basic2}) can be obtained
without using the spin tensor, that is in a quantum field theory where the stress-energy
tensor is the symmetrized Belinfante tensor. The issue is a relevant one for  
if (\ref{basic2}) was dependent on a spin tensor, $\wspt$ would acquire a full physical
meaning and relativistic hydrodynamics - to say the least - should be extended 
\cite{florkowski} to fluids with spin.  

Another major theoretical problem is to find an exact solution for the polarization
at global thermodynamic equilibrium with $\varpi = {\rm const} \ne 0$, that is when
the (\ref{gencov}) becomes:
\be\label{densop2}
  \rho = \frac{1}{Z} \exp \left[ - b_\mu {\wP}^\mu  
  + \frac{1}{2} \varpi_{\mu\nu} \wJ^{\mu\nu} + \zeta \wQ \right]
\ee
Indeed, the formulae (\ref{basic}),(\ref{basic2}) are first-order expansions in
thermal vorticity and based on an educated {\em ansatz} on the Wigner function of the 
Dirac field with the density operator (\ref{densop2}). Even though the formula 
(\ref{basic}) appears to be the only reasonable first-order expression, finding the 
exact solution would certainly be a benchmark in the theory of relativistic fluids
with polarization and a crucial tool to settle open issues.

\section*{Acknowledgments}

I am greatly indebted with I. Karpenko for his invaluable help in preparing this
talk. The author would like to thank the organizers for their excellent work in arranging
a very enjoyable conference and for their patience in waiting for my contribution
to the proceedings.


\end{document}